\documentclass[twocolumn,5p,authoryear,final]{elsarticle4arXiv}
\usepackage{framed,multirow}

\usepackage{amssymb}
\usepackage{latexsym}

\usepackage[switch]{lineno}

\usepackage{url}
\usepackage{xcolor}
\definecolor{newcolor}{rgb}{.8,.349,.1}

\usepackage[citebordercolor=white,
    colorlinks=true,
    linkcolor=blue,                 % internal links
    citecolor=blue,                 % Links for references
    urlcolor=red,                   % Links for URLS
    ]{hyperref}

\usepackage{booktabs} % \toprule etc.
\usepackage{siunitx}
\usepackage{threeparttable}

\def\wrt{\mbox{w.r.t. }}
\newcommand{\bis}[2]{#1\,--\,#2}

\definecolor{blau}{rgb}{0.9,0.9,1.0}
\definecolor{rot}{rgb}{1.0,0.8,0.8}
\definecolor{grun}{rgb}{0.0,0.5,0.0}

\defcitealias{RINEX_4}{IGS RWG and RTCM, 2021} 	

\journal{Advances in Space Research}

\begin{document}

    \begin{frontmatter}
        
        \title{BeiDou-3 orbit and clock quality of the IGS Multi-GNSS Pilot Project}
          
        \author[1]{Peter Steigenberger\corref{cor1}}
        \cortext[cor1]{Corresponding author: peter.steigenberger@dlr.de}
        \author[2]{Zhiguo Deng}  
        \author[3]{Jing Guo}
        \author[5]{Lars Prange}
        \author[6]{Shuli Song}
        \author[1]{Oliver Montenbruck}
        
        \address[1]{German Space Operations Center (GSOC), Deutsches Zentrum f\"ur Luft-und Raumfahrt (DLR),\\ M\"unchner Stra\ss e 21, D-82234 We\ss ling, Germany}
        
        \address[2]{Helmholtz-Zentrum Potsdam -- Deutsches GeoForschungsZentrum, Telegrafenberg, D-14473 Potsdam, Germany}      
        
        \address[3]{GNSS Research Center, Wuhan University, No. 129 Luoyu Road, Wuhan 430079, China}      
                
        \address[5]{Astronomical Institute, University of Bern, Sidlerstrasse 5, CH-3012 Bern, Switzerland}
        
        \address[6]{Shanghai Astronomical Observatory, Chinese Academy of Sciences, 80 Nandan Road, Shanghai 200030, China}

\begin{abstract}
Within the Multi-GNSS Pilot Project (MGEX) of the International GNSS Service (IGS), precise orbit and clock products for the BeiDou-3 global navigation satellite system (\mbox{BDS-3}) are routinely generated by a total of five analysis centers. The processing standards and specific properties of the individual products are reviewed and the \mbox{BDS-3} orbit and clock product performance is assessed through direct inter-comparison, satellite laser ranging (SLR) residuals, clock stability analysis, and precise point positioning solutions. The orbit consistency evaluated by the signal-in-space range error is on the level of \SIrange{4}{8}{cm} for the medium Earth orbit satellites whereas SLR residuals have RMS values between 3 and \SI{9}{cm}. The clock analysis reveals sytematic effects related to the elevation of the Sun above the orbital plane for all ACs pointing to deficiencies in solar radiation pressure modeling. Nevertheless, precise point positioning with the \mbox{BDS-3} MGEX orbit and clock products results in 3D RMS values between 7 and \SI{8}{mm}. 
\end{abstract}

        
\end{frontmatter}

%% For linenumbers
%\linenumbers
\vspace{1cm}
\section{Introduction}                                      \label{sec:intro}
%%%%%%%%%%%%%%%%%%%%%%%%%%%%%%%%%%%%%%%%%%%%%%%%%%%%%%%%%%%%%%%%%%%%%%%%%%%%%

The Chinese BeiDou-3 system  \citep[BDS-3,][]{Yang_2019a} is one of currently four Global Navigation Satellite Systems (GNSSs). With the first satellite launched in 2017, it was declared operational in July 2020\footnote{\url{http://en.beidou.gov.cn/WHATSNEWS/202008/t20200803_21013.html}}.
BDS-3 is the global extension of the regional BeiDou-2 system built up between 2009 and 2012 and operational since December 2012. As of the beginning of 2022, the BDS-3 constellation consists of 24 satellites in medium Earth orbit (MEO), 3 satellites in inclined geosyncronous orbit (IGSO) and 3 satellites in geostationary Earth orbit (GEO)\footnote{\url{http://www.csno-tarc.cn/en/system/constellation}}. The BeiDou-3 test satellites launched in 2015 and 2016 are denoted as BDS-3S. These two MEO and two IGSO satellites are not part of the operational constellation and have currently the status \textit{Experiment}.

The Multi-GNSS Pilot Project \citep[MGEX;][]{Montenbruck_2017} of the International GNSS Service \citep[IGS;][]{IGS_2017} aims at ``a comprehensive integration of multi-GNSS tracking and analysis into all IGS components and activities''. It is the successor of the Multi-GNSS Experiment established in 2011. Whereas the multi-GNSS tracking network is fully integrated in the IGS since 2016, dedicated multi-GNSS orbit and clock products are generated by currently seven MGEX analysis centers (ACs). Five of these ACs include BDS-3 in their products, see Table~\ref{tab:MGEXAC}. The quality of Galileo MGEX orbit and clock products was assessed by \cite{Steigenberger_2015}. The more recent analysis of \cite{Steigenberger_2019} covers GPS, GLONASS, Galileo, and BDS-2 but lacks BDS-3 as only one AC considered this system at that time.

% Not considered: \cite{Yan_2018} BDS-3 
%  PCOs \cite{Xia_2020b},

Initial results of \mbox{BDS-3S} precise orbit determination (POD) are given in \cite{Xie_2017}  and  \cite{Zhao_2018}. They report a radial orbit accuracy evaluated by satellite laser ranging (SLR) of \SIrange{10}{30}{cm} and \SI{10}{cm}, respectively. \cite{Yan_2019a} developed an a priori box-wing model for the \mbox{BDS-3} MEO satellites. Compared to a purely empirical orbit modeling, they could achieve an improvement in the orbit overlap RMS of about \SIrange{10}{20}{\percent} and SLR residuals of about \SI{5}{cm}.

Availability of satellite metadata is a prerequisite for proper modeling of satellite orbits.  \cite{SECM_2018} and \cite{CSNO_2019a} published the satellite mass, approximate dimensions, retroreflector offsets, phase center offsets, attitude law, and a subset of the optical properties. \cite{Li_2020} utilized these metadata and could demonstrate improvements in the orbit overlap RMS by up to \SI{1}{cm} when applying appropriate models for attitude and a box-wing model for the solar radiation pressure based on the surface areas and properties. The \mbox{BDS-3} MEO orbits computed with these sophisticated models entail SLR residuals of \SIrange{3}{6}{cm}. \cite{Xu_2019a} report a similar radial orbit accuracy of \SIrange{4}{6}{cm} with a purely empirical modeling of the solar radiation pressure. Orbit determination based on inter-satellite links (ISLs) was studied by several authors \citep{Wang_2019f,Ruan_2020b,Guo_2020b,Xie_2020b} but none of the MGEX ACs uses ISL data due to lack of public availability.

\cite{Chen_2021e} found that the clock stability of the \mbox{BDS-3} satellites is improved compared to BDS-2 but also shows periodicities at the orbital frequencies. \cite{Cao_2021b} confirm the improved quality of the observed \mbox{BDS-3} clocks although the Galileo clocks slightly outperform them. They report an average clock RMS of \SI{0.15}{ns} \wrt a second order polynomial for a reduced set of BeiDou-3 MEO satellites. The latter results were also confirmed by \cite{Sun_2020b}. \cite{Gu_2021b} report a slightly better performance of the GFZ MGEX \mbox{BDS-3} clock products compared to WUM evaluated by Allan deviations. 

% ADEV analysis: \cite{Qin_2020}
% BDS-2 only: \cite{Wang_2021}
%NEW: \cite{Geng_2022b}

\begin{table*}[ht!]
    \centering
    \begin{threeparttable}
        
        \caption{Analysis centers contributing BeiDou-3 orbit products to IGS MGEX. The product ID is composed of a 3-character AC code, a 1-digit version number (\texttt{0} for all MGEX products), a 3-character project code (\texttt{MGX} for MGEX) and a 3-dcharacter solution code (\texttt{RAP} for rapid and \texttt{FIN} for final). }
        \label{tab:MGEXAC}
        \begin{small}

        \begin{tabular}{l l l c c c l}\toprule
            Abb. & Institution, Country                      & Product ID          & MEO & IGSO & GEO & Reference\\\midrule
            CODE & Center for Orbit Determination in Europe, Switzerland  & \texttt{COD0MGXFIN} & x   & x    &     & \cite{Prange_2020b} \\
            GFZ  & Deutsches GeoForschungsZentrum, Germany   & \texttt{GFZ0MGXRAP} & x   & x    & x   & \cite{Deng_2017} \\
            IAC  & Information and Analysis Center, Russia   & \texttt{IAC0MGXFIN} & x   & x    & x\tnote{1}     & \\ 
            SHAO & Shanghai Astronomical Observatory, China  & \texttt{SHA0MGXRAP} & x   & x    & x\tnote{2}    & \\       
            WU   & Wuhan University, China                   & \texttt{WUM0MGXFIN} & x   & x    & (x)\tnote{3}    & \cite{Guo_2016c}\\ \bottomrule
        \end{tabular}
        \begin{tablenotes}\footnotesize 
            \item[1] GEO-1 very sparse
            \item[2] only GEO-1
            \item[3] GEO-1 included for about 40 days end of 2019 and \bis{279}{281}/2020
        \end{tablenotes}
    \end{small}
    \end{threeparttable}
\end{table*}

As of early 2022, the IGS tracking network comprises 514 stations, 308 of them are able to track BeiDou. However, not all of them are suitable for \mbox{BDS-3} orbit and clock estimation due to single-frequency tracking for \mbox{BDS-3} or limitations regarding the supported pseudo-random noise (PRN) numbers. These restrictions are related to either outdated receiver hardware, up-to-date receivers with outdated firmware, or individual user settings. Early firmware versions were limted to PRNs up to C37 but improved versions supporting PRNs up to C63 were made available in 2019 by different receiver manufacturers \citep{Steigenberger_2020}. Regarding hardware, the Leica GR10 and GR25 as well as the Trimble NetR9 receivers are limited to PRNs up to C30. This means that for the 64 BDS-capable IGS stations equipped with these receiver types, only 12 out of the 24 \mbox{BDS-3} MEOs are tracked and none of the IGSO and GEO satellites. 

In this paper, satellites are uniquely identified by their space vehicle number (SVN) composed by a character representing the constellation  (C for BeiDou) and a 3-digit number. The third \mbox{BDS-3} GEO satellite (SVN C230) launched in June 2020 is not yet set healthy for the precise navigation and timing (PNT) services and has the status \textit{Testing}. All other \mbox{BDS-3} satellites are part of the operational constellation. \mbox{BDS-3} MEO satellites are provided by two different manufacturers: the China Academy of Space Technology (CAST, 14 satellites) and the Shanghai Engineering Center for Microsatellites (SECM, 10 satellites). \mbox{BDS-3} IGSO and GEO satellites were all manufactured by CAST. The latest two SECM spacecraft (C225 and C226) have an updated satellite bus with a slightly larger surface area.

Section~\ref{sec:products} introduces the different modeling options of the MGEX ACs. The orbit and clock quality are assessed in Sect.~\ref{sec:orbit} and \ref{sec:clock}, respectively, by orbit comparisons, satellite laser ranging residuals, clock RMS values, and Allan deviations. Section~\ref{sec:ppp} evaluates the combined orbit and clock quality by analyzing the precise point positioning performance of the individual AC orbit and clock products.

%%%%%%%%%%%%%%%%%%%%%%%%%%%%%%%%%%%%%%%%%%%%%%%%%%%%%%%%%%%%%%%%%%%%%%%%%%%%%
\section{IGS MGEX BDS-3 orbit and clock products}        \label{sec:products}
%%%%%%%%%%%%%%%%%%%%%%%%%%%%%%%%%%%%%%%%%%%%%%%%%%%%%%%%%%%%%%%%%%%%%%%%%%%%%
% SHAO from Li et al. (2019)
\begin{table*}
 \begin{threeparttable}
  \begin{footnotesize}
  \caption{BDS-3 modeling options and estimated parameters of the IGS MGEX analysis centers considered in this study. Abbreviations: A  -- along-track; C -- cross-track; ISB -- inter-system bias; PCOs -- phase center offsets; PCVs -- phase center variations; PO2 -- 2nd order polynomial; PS -- pseudo-stochastic; PWL -- piece-wise linear; R -- radial; Sc -- scaling factor; $\beta$ -- elevation of the Sun above the orbital plane. C2I and C6I stand for the in-phase component of the B1 and B3 pseudorange signals according to the RINEX observation codes \citepalias{RINEX_4}} \label{tab:AC_opt}
  \centering
  \begin{tabular}{llllll}\toprule
                 &  CODE      & GFZ	          &  IAC  & SHAO  & WU \\ \midrule                
      
      Software	 &  Bernese GNSS  & EPOS.P8       & STARK & SPADA \tnote{a} & PANDA  \\
                 &  Software 5.3  &               &  & &   \\
      
      Stations \tnote{b}   & $\approx$140 ($\approx$90)  & \bis{150}{160} & $\approx$110 ($\approx$80) &    $\approx$130         &  $\approx$140  ($\approx$110)   \\

      Differencing & Double diff. (orbit)  & Undifferenced & Undifferenced  & Undifferenced & Undifferenced  \\
                   & Undifferenced (clock) &               &   &   &           \\
      
      Data interval & 72 h (orbit)  &      24 h & 48 h  & 24 h & 24 h \\
                    & 24 h (clock)  &           &       &  &      \\
      
      Data sampling & 3 min (orbit)  &  5 min (orbit)  & 5 min (orbit)  &   5 min     & 5 min. (orbit)     \\
                   & 5 min (clock) \tnote{c} &   30 s (clock) &  30 s (clock)   &      & 30 s (clock) \tnote{d}           \\
      
      BDS-3 ref. sig.  & C2I/C6I & C2I/C6I & C2I/C6I  & C2I/C6I &  C2I/C6I     \\
            
      Elevation cutoff & 3$^\circ$ (orbit)     & 7$^\circ$	  &  7$^\circ$    &  7$^\circ$  & 7$^\circ$ (GPS,GLO)     \\
      & 5$^\circ$ (clock)     &               &      &    &10$^\circ$ (GAL, BDS, QZS) \\
      
      Elevation-depen- &  $w=\sin \epsilon$ (phase) &  $w= 2 \sin \epsilon$ for $\epsilon < $\SI{30}{\degree} &   $w= 2 \sin \epsilon$ for $\epsilon < $\SI{30}{\degree} &  $w= 2 \sin \epsilon$ for $\epsilon < $\SI{30}{\degree}  &  $w=2 \sin \epsilon$ for $\epsilon < 30^\circ$ \\ 
      
      dent weighting \tnote{e}   & $w=\sin^2 \epsilon$ (code)  &  $w=1$ for $\epsilon \geq $\SI{30}{\degree} &  $w=1$ for $\epsilon \geq $\SI{30}{\degree} & $w=1$ for $\epsilon \geq $\SI{30}{\degree}   &  $w=1$ for $\epsilon \geq $\SI{30}{\degree}  \\
      
      Rec. ant. model       & igsR3.atx &  igs14.atx      & igs14.atx &   igs14.atx         &  igs14.atx   \\      
      
      Sat. ant. model       & \cite{CSNO_2019a}   &  igs14.atx      & igs14.atx &  \cite{CSNO_2019a}  &  igs14.atx   \\
      
            \midrule
      \multicolumn{6}{c}{BDS-3 specific options} \\
      \midrule

%      Sat. ant. PCVs         & not applied  &  not applied & not applied         &           & not applied    \\
            
      CAST attitude    & \cite{Dilssner_2017} & \cite{Wang_2018b} & \cite{Dilssner_2017} & \cite{Dilssner_2017} &  \cite{Wang_2018b}   \\
      
      SECM attitude    & \cite{Dilssner_2017} & \cite{Zhao_2018}  & \cite{Dilssner_2017} & \cite{Dilssner_2017} & \cite{CSNO_2019a}    \\ 
      
      A priori SRP     & none       &  none   & box-wing& box-wing &  box-wing   \\  
                       &            &         & \citep{CSNO_2019a}  &  \citep{CSNO_2019a} & \citep{Wang_2019g}  \\
      Antenna thrust \tnote{f}    &  \SI{310}{W}/\SI{280}{W}/\SI{0}{W}         &      \SI{310}{W}/\SI{280}{W}/\SI{100}{W}       &   \SI{200}{W}/\SI{200}{W}/\SI{200}{W}      &   --        & \SI{310}{W}/\SI{280}{W}/\SI{100}{W}   \\
      
      Earth albedo/IR  & not applied & not applied & \cite{Yan_2019a} & not applied & applied    \\
      
      \midrule
      \multicolumn{6}{c}{Estimated parameters} \\
      \midrule
      Station coord.	   & 72 h    & 24 h & 48 h  &  24 h  &  24 h \\
      
      Trop. zen. delay & 2 h PWL  & 1 h random-walk & 3 h PO2 &  1 h PWL & 2 h \\
      
      Trop. gradients  & 24 h PWL & 2 h random-walk  & 48 h   & 12 h PWL & 24 h  \\
      
      Receiver clocks  & 5 min	 & 30~s/5 min \tnote{g}	  & 30 s & 5 min & 30 s \tnote{h}   \\
      
      GPS ambiguities  & fixed \tnote{i}     & fixed    & fixed & fixed & fixed \tnote{j}  \\
      
      BDS-3 ambig.     & fixed \tnote{k}     & fixed    & float & fixed & fixed \tnote{j}  \\
      
      BDS-3 clocks     & 30 s \tnote{c}     &  30 s  & 30 s & 5 min    &   30 s \\
      
      BDS-3 SRP        & ECOM-2       & ECOM     & ECOM-based & ECOM  &  ECOM    \\
                           & D0, Y0, B0, BC, BS, & D0, Y0, B0, BC, BS &  Sc, Y0, BC, BS &  D0, Y0, B0, BC, BS &   D0, Y0,  B0, BC, BS   \\
                           & D2C, D2S  &       &$|\beta|<$\SI{5}{\degree} D0, B0 &  & + constant along-track  \\
                       &            &           & $|\beta|>$\SI{5}{\degree} D1C, D1S       &  & \\
                           
      PS pulses & 12 h in R, A, C &  at noon in  R, A, C \tnote{l}  & at noon in  R, A, C  & -- & -- \\                     
      
      Bias parameters    & 24 h ISB  & 24 h ISB   & 48 h ISB  & 24 h ISB  &     24 h ISB     \\
      \bottomrule
  \end{tabular}
  \begin{tablenotes}\footnotesize
      \item[a] the SPAce-geodetic Data Analysis (SPADA) software package is developed based on PANDA provided by the GFZ real-time GNSS group
      \item[b] the number in brackets is the number of BDS-capable stations   
      \item[c] \SI{30}{s} clock estimates are obtained from \SI{5}{min} clocks according to \cite{Bock_2009}
      \item[d] GPS and GLONASS \SI{30}{s} clock estimates are obtained from \SI{5}{min} clocks by interpolation based on epoch-differenced carrier phase measurements; BDS-3 \SI{30}{s} clock parameters are estimated directly
      \item[e] the weighting factor $w$ describes the variation of the measurement standard deviation $\sigma\left(\epsilon \right) = \sigma_0 / w\left(\epsilon\right)$ with elevation $\epsilon$ 

      \item[f] MEO CAST/MEO SECM/IGSO

      \item[g] \SI{30}{s} sampling for reduced set of selected stations with highly stable clocks
      \item[h] fixed to PPP for BDS orbit determination

      \item[i] widelane/narrowlane approach for double-difference and zero-difference solution
      \item[j] \cite{Ge_2005}
      \item[k] widelane/narrowlane approach for double-difference solution, widelane approach for zero-difference solution
      \item[l] only for MEO satellites   
  \end{tablenotes}
      
  \end{footnotesize}
 \end{threeparttable}
\end{table*}

The five MGEX ACs providing \mbox{BDS-3} products are listed in Table~\ref{tab:MGEXAC}. Deutsches GeoForschungsZentrum (GFZ) and Shanghai Astronomical Observatory (SHAO) provide rapid products with a latency between one and three days whereas the other ACs contribute final orbit and clock products with latencies between four days and three weeks. The Center for Determination in Europe (CODE) and the Information and Analysis Center (IAC) publish their products in weekly batches, the other ACs have a daily update cycle. The products are available at the global data centers of the IGS, e.g., the Crustal Dynamics Data Information System \citep[CDDIS,][]{Noll_2010}. Wuhan University (WU) was the first MGEX AC that started processing \mbox{BDS-3} on January 1, 2019. Due to limited availability of tracking data \citep{Steigenberger_2020}, only satellites up to PRN number C37 were considered. WU started to include the full range of MEO PRNs on day of year (DoY) 279/2020. CODE added \mbox{BDS-3} MEO and IGSO satellites to their final MGEX products on DoY 66/2021. 

In the GFZ rapid products, \mbox{BDS-3} is included since 166/2020. Products of IAC located near Moscow, Russia, are available since September 12, 2020 (DoY 256/2020) considering \mbox{BDS-3} from the very beginning. SHAO processes \mbox{BDS-2} and \mbox{BDS-3} starting with DoY 148/2020. Whereas all ACs process the MEO and IGSO satellites, the coverage of the GEO satellites is sparse. The unhealthy \mbox{BDS-3} GEO-3 satellite is not considered by any AC due to insufficient observation data provided by the IGS tracking network. Furthermore, the experimental BDS-3S satellites are also excluded from the analysis because of their limited relevance for common users and scientific applications.

Table~\ref{tab:AC_opt} summarizes important modeling options and details about estimated parameters of the five MGEX ACs. CODE estimates the satellite orbits in a double-difference approach and fixes them when solving for \SI{5}{min} clock parameters in undifferenced mode. Based on this clock solution, \SI{30}{s} high-rate clocks are generated according to \cite{Bock_2009}. The other ACs use undifferenced observations in all processing steps and similar approaches for clock densification.

WU uses a three-step approach for their MGEX contribution. In the first step, satellite orbits and clocks as well as Earth rotation paramters (ERPs) are estimated from GPS and GLONASS observations. The GPS estimates of the first step are fixed in the second step solving for station coordinates, troposphere zenith delays, and receiver clocks in a precise point positioning (PPP). The third step comprises BDS-3 orbit and clock estimation keeping the PPP results fixed. 

The number of processed stations varies between 110 and 160 stations for the different ACs. However, the number of stations with BDS tracking capability is smaller, roughly two-thirds. Tracking data for PRNs $>$ C37 is still limited. E.g., for the IAC solution, only 65 compared to 90 stations provide observations for PRNs $>$ C37.

All ACs use the BDS B1 and B3 frequency bands for their BDS-3 solutions due to overlap with the BDS-2 signals and better station coverage compared to the BDS-3-only signals in the B1C, B2a, and B2b bands. Although not considered in our analysis, all ACs include BDS-2 in their orbit and clock products. BDS-2 and BDS-3 are usually treated as one system, i.e., only one inter-system bias (ISB) parameter per station \wrt GPS is estimated. \cite{Jiao_2022} showed that a separate treatment of \mbox{BDS-2} and \mbox{BDS-3} can improve the overall product consistency. Therefore, GFZ considers a separate ISB estimation for 2nd and 3rd generation BDS starting with 161/2021.

Individual satellite antenna phase center offsets (PCOs) of all BDS-3 satellites are published by \cite{CSNO_2019a} but the IGS antenna model igs14.atx \citep{Rebischung_2016b} currently still contains block-specific values. Receiver antenna calibrations in igs14.atx only cover the legacy frequency bands L1 (\SI{1575.42}{MHz}) and L2 (\SI{1227.60}{MHz}). Therefore, ACs using igs14.atx substitute antenna parameters for the \mbox{BDS-3} frequency bands B1 (\SI{1561.098}{MHz}) and B3 (\SI{1268.52}{MHz}) by the respective L1 and L2 values. In order to avoid this slight mismodeling, CODE applies the anntena model igsR3.atx compiled for the 3rd reprocessing effort of the IGS and complemented by the CSNO BDS-3 satellite antenna PCOs \citep{Villiger_2021}. igsR3.atx includes multi-GNSS receiver antenna calibrations for 37 antennas also covering the B1 and B3 bands \citep{Wuebbena_2019,Zimmermann_2019}. Boresight-dependent group delay variations that are known to affect \mbox{BDS-2} \citep{Wanninger_2014} are essentially negligible for \mbox{BDS-3} \citep{Beer_2021b} and therefore neglected by all ACs.

GNSS satellites in non-geostationary orbit generally apply a yaw-steering attitude pointing the navigation signal transmit antenna towards the center of the Earth and the solar panel axis perpendicular towards the Sun. However, for small elevations of the Sun above the orbital plane ($\beta$-angle), the yaw rates required to maintain this attitude mode exceed the maximum yaw rates of the attitude control system. Therefore, orbit-normal or rate-limited attitude modes have been implemented on the different types of satellites. The \mbox{BDS-3} CAST MEO satellites follow the same attitude mode as the latest BDS-2 satellites \citep{Dilssner_2017}: for $|\beta|<$\SI{2.8}{\degree}, a smoothed yaw steering is applied avoiding excessive yaw rates \citep{Wang_2018b}. The SECM BDS-3 MEO satellites keep the $\beta$-angle of \SI{\pm 3}{\degree} fixed for the yaw rate computation for $|\beta|<$\SI{3.0}{\degree} \citep{SECM_2018}. The IGSO satellites also operate in smoothed yaw steering as the CAST MEO satellites. The GEO satellites permanently operate in orbit-normal mode where the solar panel axis is perpendicular to the orbital plane. 

Appropriate modeling of the solar radiation pressure (SRP) is a key issue for generating high-precision orbit and clock products. The Empirical CODE Orbit Model \citep[ECOM,][]{Beutler_1994} as well as its successor ECOM-2 \citep{Arnold_2015} model SRP in a Sun-oriented system with D pointing towards the Sun, Y along the solar panel axis and B perpendicular to D and Y. Constant terms are indicated by the index 0, periodic terms by the indices C and S. The 5-parameter ECOM (D0, Y0, B0, BC, BS) is not able to properly model the SRP for elongated satellite bodies like Galileo \citep{Montenbruck_2015} or BDS-3. Therefore, either ECOM-2 or an a priori box-wing model should be applied. Even though basic information on the satellite geometry has been published as part of the satellite metadata by \cite{CSNO_2019a}, only incomplete optical properties have been supplied. As such, no a priori box-wing model is presently used by GFZ, which partly limits the quality of the resulting orbits. To cope with these limitations, WU makes use of an empirically adjusted BW model \citep{Wang_2019g}.

Selected \mbox{BDS-3} satellites have geometric features that are not covered by a simple box-wing model. Such features include the Search and Rescue (SAR) antenna on a limited number of MEO satellites \citep{Duan_2021d} as well as communication antennas of the IGSO and MEO satellites. None of the MGEX ACs currently considers these extended structures in their SRP modeling. Even box-wing models can only provide a first-order approximation of the actual radiation pressure. Therefore, empirical SRP parameters are usually estimated on top of the box-wing model. IAC uses a different SRP parameter setup for small $\beta$-angles and in addition estimates a scaling factor (Sc) for the accelerations of the a priori box-wing model. Another approach to compensate orbit modeling errors is the estimation of pseudo-stochastic pulses, i.e., small velocity changes. E.g., CODE estimates such pulses every \SI{12}{h} in radial (R), along-track (A), and cross-track (C) direction.

Antenna thrust is a small force acting mainly in radial direction due to the transmission of navigation signals \citep{Steigenberger_2018}. It is directly related to the transmit power and to the inverse of the satellite mass. For \mbox{BDS-3}, no transmit antenna gain pattern are available. Therefore, BDS-2 gain pattern were used for a preliminary BDS-3 MEO transmit power estimation from measurements of the \SI{30}{m} high-gain antenna of the German Space Operations Center (GSOC), located in Weilheim, Germany. The block-specific transmit power derived in this analysis amounts to \SI{310}{W} for CAST and \SI{280}{W} for SECM MEO satellites. These values are included in the IGS satellite metadata file \citep{IGS_MSX}. Due to the low elevation at Weilheim, no transmit power measurements are available for the BDS-3 IGSO and GEO satellites. 

Earth albedo and infrared (IR) radiation \citep{Rodriguez-Solano_2012b} also mainly affect the radial direction and summarize the effects of reflected and reemitted radiation from the Earth. Whereas considering these effects for GPS satellites is well established, not all ACs considers them for BDS-3.

%%%%%%%%%%%%%%%%%%%%%%%%%%%%%%%%%%%%%%%%%%%%%%%%%%%%%%%%%%%%%%%%%%%%%%%%%%%%%
\section{Orbit quality}                                     \label{sec:orbit}
%%%%%%%%%%%%%%%%%%%%%%%%%%%%%%%%%%%%%%%%%%%%%%%%%%%%%%%%%%%%%%%%%%%%%%%%%%%%%
The orbit precision is evaluated by the orbit signal-in-space range error (SISRE) computed between pairs of solutions whereas satellite laser ranging residuals allow for an assessment of the orbit accuracy. The analysis interval is limited to a time period covered by all ACs, namely day of year 66\,--\,365/2021.

\begin{table}[ht!]
    \centering
    \begin{footnotesize}
        \caption{Orbit SISRE (orb) and overall SISRE (ov) for different types of satellites. All values are given in \SI{}{cm}.}
        \label{tab:sisre}
        \begin{tabular}{llrrrrrrrr} \toprule
            AC1 & AC2 & \multicolumn{2}{c}{MEO} && \multicolumn{2}{c}{IGSO} &&\multicolumn{2}{c}{GEO}  \\ \cmidrule{3-4} \cmidrule{6-7} \cmidrule{9-10} 
            &     & orb & ov && orb & ov && orb & ov  \\\midrule
            COD  & GFZ  &  4.1 & 3.4 && 20.8 & 18.1 &&  --  & --       \\
            COD  & IAC  &  4.1 & 3.5 && 16.9 & 12.5 &&  --  & --       \\
            COD  & SHA  &  6.0 & 5.3 && 22.4 & 19.4 &&  --  & --       \\
            COD  & WUM  &  4.7 & 4.3 && 23.7 & 20.4 &&  --  & --       \\
            
            GFZ  & IAC  &  4.4 & 4.7 && 15.5 & 12.0 && 59.1 & 58.4 \\
            GFZ  & SHA  &  6.0 & 5.6 && 11.4 &  9.8 && 55.8 & 55.8 \\
            GFZ  & WUM  &  5.0 & 5.1 && 10.0 &  7.6 &&  --  & --    \\
            
            IAC  & SHA  &  7.2 & 7.0 && 18.1 & 14.8 && 54.1 & 54.1  \\
            IAC  & WUM  &  4.3 & 4.3 && 18.3 & 14.6 && --   & --     \\
            
            SHA  & WUM  &  7.6 & 7.2 && 11.4 &  8.9 && --   & --    \\
            \bottomrule
        \end{tabular}
    \end{footnotesize}
\end{table}

%----------------------------------------------------------------------------
\subsection{Orbit precision}                         \label{sec:orbPrecision}
%----------------------------------------------------------------------------
The signal-in-space range error (SISRE) is a well-established measure for the quality of broadcast orbits and clocks \citep{Montenbruck_2018}. In the following, the orbit-only SISRE
\begin{equation}
    \text{SISRE}_{\text{orb}}=\sqrt{w_1^2 R^2 +w_2^2 \left(A^2+C^2 \right)} 
\end{equation}
is used for the assessment of the consistency of the MGEX orbit products. R, A, and C represent the radial, along-track, and cross-track orbit differences and $w_1$ and $w_2$ are weighting factors depending on the orbit type \citep{Montenbruck_2018}. Gross outliers exceeding  \SI{1.0}{m} were rejected. Table~\ref{tab:sisre} lists the SISRE$_{\text{orb}}$ for MEO, IGSO, and GEO satellites included in each pair of ACs.

For the MEO satellites, the smallest orbit SISRE of \SI{4.1}{cm} is achieved for COD/GFZ and COD/IAC, and the consistency of GFZ/IAC, IAC/WUM and COD/WUM is only a few millimeters worse. The largest SISRE$_{\text{orb}}$ values of up to \SI{8}{cm} occur for the comparisons with SHA. They are mainly attributed to increased systematic along-track differences of SHAO \wrt the other ACs with a standard deviation of up \SI{15}{cm} contributing to the SISRE$_{\text{orb}}$ with a scaling factor of $w_2^2 \approx \frac{1}{7}$. \cite{Steigenberger_2019} reported orbit SISRE values of \SIrange{7}{14}{cm} for the BeiDou-2 MEO orbits of COD, GFZ, and WUM for the time period January until June 2018. The current BDS-3 consistency of these ACs outperforms that result by a factor of about two, probably driven by an increased station number, the fully populated 24-satellite MEO constellation, and improved SRP modeling. 

The orbit-only SISRE of the IGSO satellites is significanty larger compared to the MEO satellites and ranges from 10 to \SI{24}{cm}. This can be attributed to the limited, regional ground station coverage of these orbits, as well as radiation pressure modeling issues related to large communication antennas \citep{Zhao_2022b} that are not covered by the published box-wing models. The best agreement is found for GFZ and WUM, the worst for COD and WUM. In contrast to the MEO satellites, the consistency of SHA with the other ACs is much better. The GEO orbit SISRE is on the level of about \SI{60}{cm}. This is mainly attributed to a smaller observation number as well as a near-static observing geometry making the decorrelation of clocks, ambiguities, and orbit parameters difficult. Further difficulties regarding SRP modeling arise from two large communication antennas \citep{Zhao_2022b} with non-disclosed shape and material properties. The overall SISRE also given in Table~\ref{tab:sisre} will be discussed in Sec.~\ref{sec:clock}.

%----------------------------------------------------------------------------
\subsection{Orbit accuracy}                           \label{sec:orbAccuracy}
%----------------------------------------------------------------------------
Other than the inter-agency comparison, which mainly provides information on the overall precision and repeatability of \mbox{BDS-3} orbit solutions, the accuracy of individual AC products can be assessed through SLR tracking. Although all \mbox{BDS-3} satellites are equipped with retro-reflector arrays (LRAs) for SLR, only a limited number is regularly observed by the tracking stations of the International Laser Ranging Service \citep[ILRS;][]{Pearlman_2002}, namely two CAST MEOs \citep[MEO-2/3, C202/6;][]{Weiguang_2018,Weiguang_2018b} and two SECM MEOs \citep[MEO-9/10, C207/8;][]{Weiguang_2018c,Weiguang_2018d}.

For the computation of SLR residuals, station coordinates are fixed to SLRF2014 \citep{ILRS_2020} and tropospheric delays are modeled according to \cite{Mendes_2004}. The LRA offset values of \cite{CSNO_2019a} and an elevation cut-off angle of \SI{10}{\degree} are applied. Outliers exceeding \SI{50}{cm} are rejected. Ocean tidal loading is considered with the FES2014b model \citep{Lyard_2020} obtained from the ocean tide loading provider of \cite{Scherneck_1991}. The number of normal points in the analysis period is about 2200 for the CAST and 1500 for the SECM satellites. 

\begin{figure}
    \centering
    \includegraphics[width=\columnwidth]{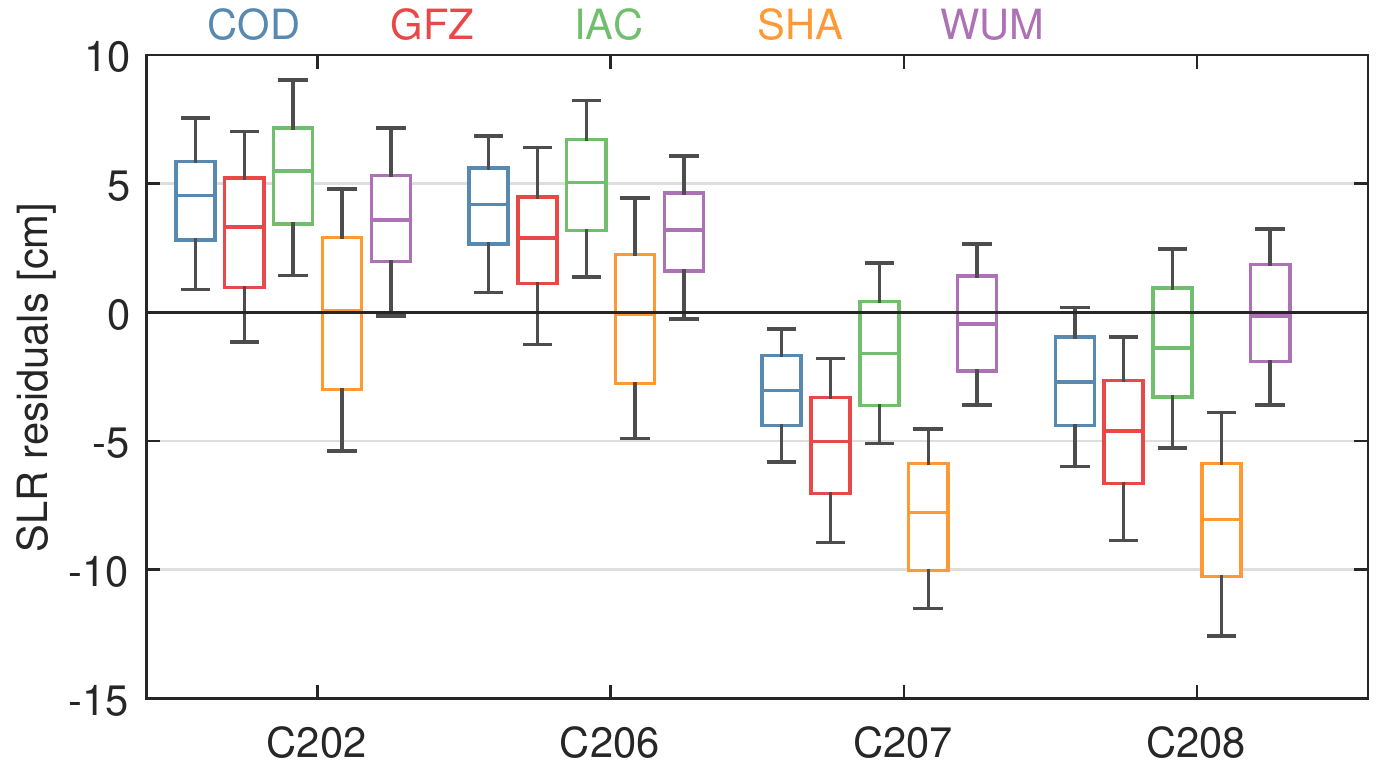} % 2 col
    \caption{SLR residuals for the time period 66\,--\,365/2021. The C202/C206 satellites are manufactured by CAST, C207/C208 by SECM. The boxes indicate the interquartile range, the whiskers the standard deviation, and the colored lines within the boxes the median value.}
    \label{fig:slrres}
\end{figure}

The distribution of SLR residuals for the four spacecraft is illustrated by box-whisker plots in Fig.~\ref{fig:slrres}. For all ACs, the satellites of the two manufacturers show significantly different biases. The biases of COD, GFZ, IAC, and WUM are in the range of \SIrange{3}{4}{cm} for the CAST satellites. The biases of the SECM satellites show a larger scatter amongst the different ACs compared to the CAST satellites. They are almost zero for WUM and vary between $-2$ and \SI{-4}{cm} for COD, GFZ, and IAC. The biases of SHA significanly differ from  the other ACs. For the CAST satellites, they are almost zero whereas they reach up to \SI{-8}{cm} for the SECM satellites.

The overall differences in the mean orbital radius and thus the scale of the different orbit products can be attributed to AC-specific modeling options like antenna thrust or albedo. In contrast to this, systematic scale differences between CAST and SECM satellites are more likely caused by SRP modeling deficiencies related to shading effects or unmodeled surface elements. As an example, consideration of the shadow cast by the LRA support plate on GIOVE-B was found to change the mean orbital radius and reduced the SLR bias by about \SI{10}{cm} \citep{Steigenberger_2015e}. More detailed spacecraft descriptions would be required, though, to assess such effects for the \mbox{BDS-3} satellites. The standard deviations of the SLR residuals are on the \SIrange{3}{4}{cm} level for COD, GFZ, IAC, and WUM and slightly higher for SHA. The RMS values as an overall measure for the orbit accuarcy as evaluated by SLR residuals vary between 4 and \SI{9}{cm}. 

The analysis of AC-specific radial orbit differences for the four BDS-3 MEO satellites tracked by SLR can be extended to the whole constellation by comparing the orbit scale differences of pairs of ACs. These scale differences have been estimated by a 7-parameter similarity transformation on a daily basis using orbit positions at \SI{5}{min} sampling. Depending on the specific pair of ACs, median values of the daily orbit scale differences exhibit a peak value of about \SI{2.1}{ppb}. This corresponds to radial orbit difference of \SI{5.9}{cm} at BDS-3 MEO altitude and is well consistent with the inter-AC radial orbit differences of the four individual satellites considered in Fig.~\ref{fig:slrres}.

\begin{figure}
    \centering
    \includegraphics[width=0.48\columnwidth]{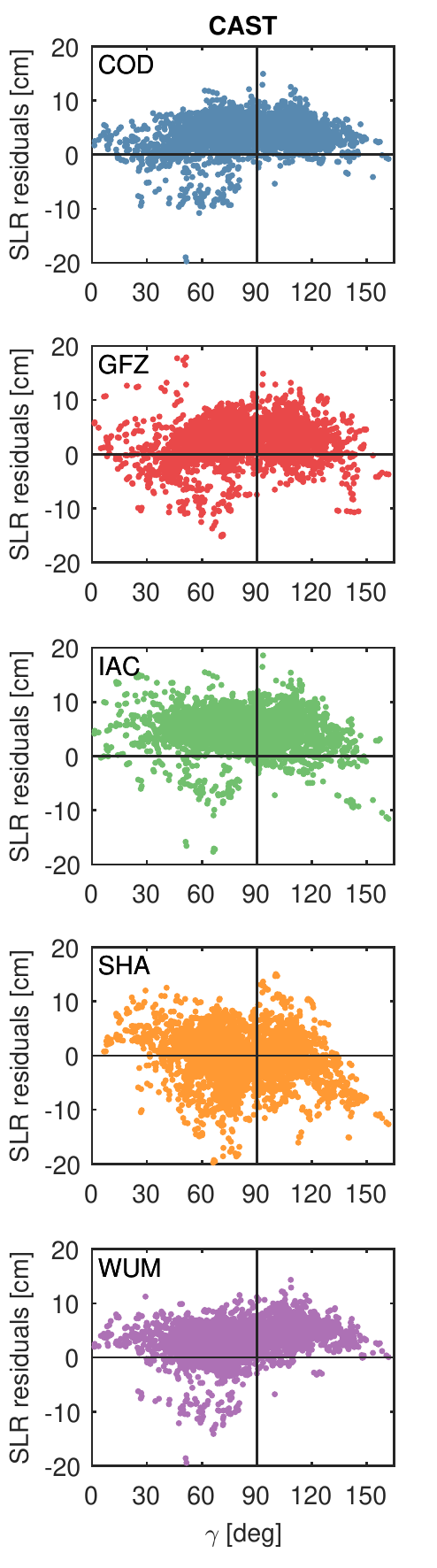} % 2 col
    \includegraphics[width=0.48\columnwidth]{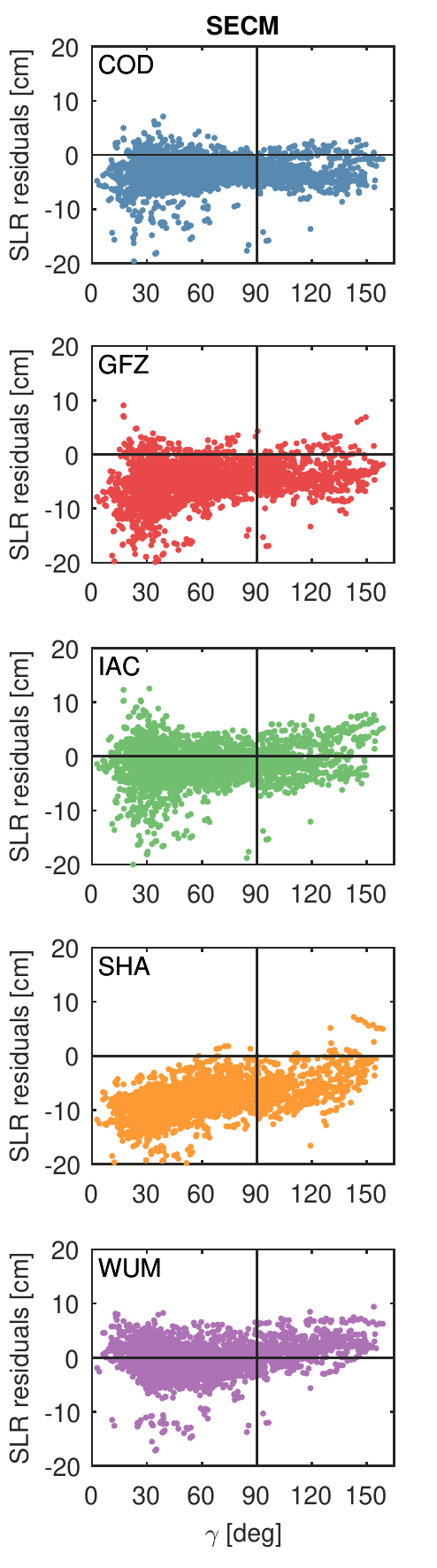}    
    \caption{SLR residuals versus Earth-satellite-Sun angle $\gamma$ for the CAST MEO satellites C202/C206 and the SECM MEO satellites C207/C208.}
    \label{fig:slrgamma}
\end{figure}

Figure~\ref{fig:slrgamma} shows the SLR residuals of the five ACs for the CAST and SECM MEO satellites plotted against the Earth-satellite-Sun angle $\gamma$. As there are no major differences for the CAST satellites C202 and C206 as well as the SECM satellites C207 and C208, the two satellites of each manufacturer are jointly shown in the respective plots. In addition to the biases already discussed above, systematic patterns can be seen for the different ACs and satellite groups. For the CAST satellites, COD, GFZ, and WUM have a positive slope of 0.022, 0.027, and \SI{0.033}{cm/\degree}, repectively. The shape of the SLR residuals is quite similar for COD and WUM althouth they use different SRP modeling approaches, namely ECOM-2 and ECOM + box-wing.

The SLR residuals of the SECM satellites show a completely different pattern compared to CAST. All ACs have a positive slope although it is almost negligible for COD with \SI{0.002}{cm/\degree}. With 0.032 and \SI{0.059}{cm/\degree}, the largest slopes are present for GFZ and SHA. The systematic patterns in the SLR residuals and the differences between CAST and SECM satellites clearly point to remaining deficiencies in the SRP modeling. Whereas the empirical \mbox{ECOM-2} used by COD achieves the smallest systematic effects, the deficiencies of a pure ECOM as applied by GFZ are obvious. But also when applying an a priori box-wing model as done by IAC, SHA, and WUM, systematic effects in the SLR residuals remain due to incomplete/inaccurate knowledge about the surface optical properties and other modeling issues.

\begin{figure*}[ht!]
    \centering
    \includegraphics[width=\textwidth]{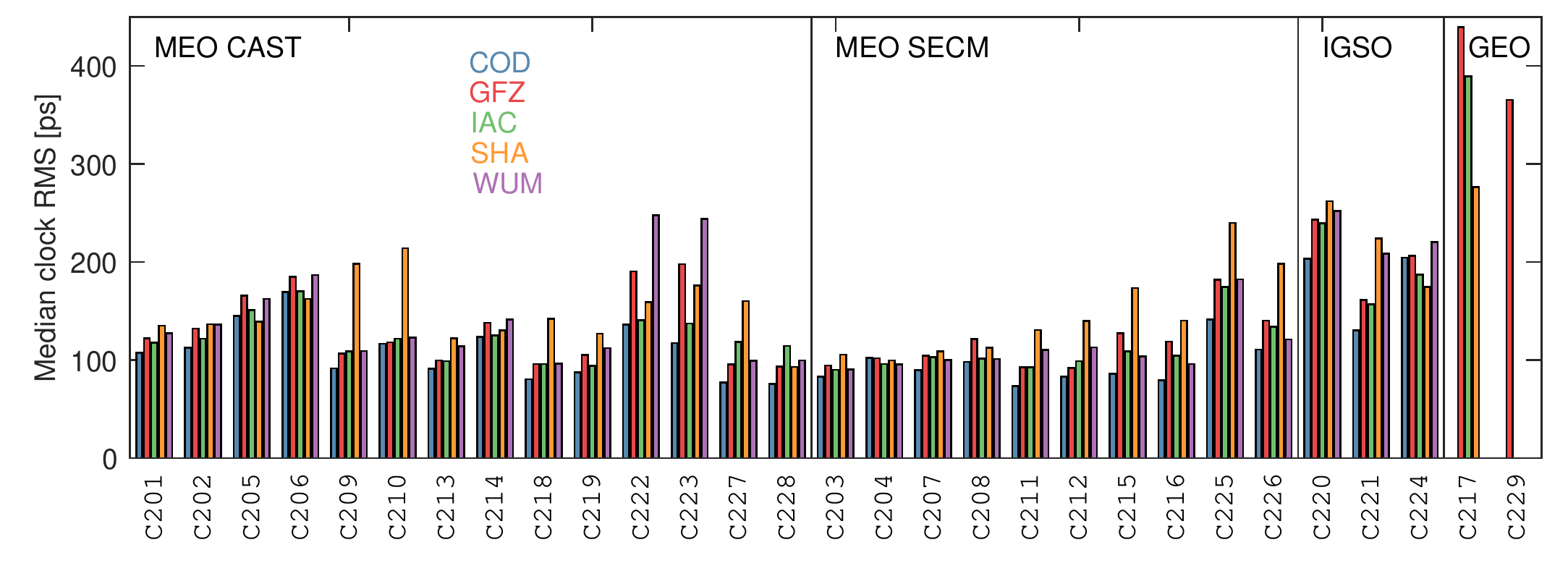}
    \caption{Median clock RMS values \wrt a daily linear fit. For SHAO, only data of November and December 2021 has been considered.}
    \label{fig:clockrms}
\end{figure*}

%%%%%%%%%%%%%%%%%%%%%%%%%%%%%%%%%%%%%%%%%%%%%%%%%%%%%%%%%%%%%%%%%%%%%%%%%%%%%
\section{Clock quality}                                   \label{sec:clock}
%%%%%%%%%%%%%%%%%%%%%%%%%%%%%%%%%%%%%%%%%%%%%%%%%%%%%%%%%%%%%%%%%%%%%%%%%%%%%
Different types of atomic frequency standards are used for the BeiDou-3 satellites. The SECM MEO satellites are equipped with two passive hydrogen masers (PHMs) as primary clocks and two rubidium atomic frequency standards (RAFSs) as secondary clocks. The CAST MEO satellites up to SVN C219 have four RAFSs onboard, later CAST MEO spacecraft have one PHM as primary clock and two RAFSs as backup clocks. The IGSO and GEO satellites are equipped with two PHMs and two RAFSs \citep[see][ and references  therein]{Zhao_2022b}.  

The currently active clock is given on the CSNO/TARC constellation status website \citep{BDS_Status} but no history is available. 
Therefore, the active clock as of September 9, 2021, as given in \cite{Zhao_2022b} is assumed to be the active clock of the whole analysis interval. This assumption seems to be justified for all satellites except for C220 showing a significant reduction in RMS as well as improved Allan daviation for all ACs in June 2021. 

Clock estimates obtained from GNSS pseudorange and phase observations are affected by orbit modeling deficiencies due to correlations between the radial orbit errors and the satellite clock parameters. \cite{Xie_2020b} demonstrated the presence of SRP-induced apparent clock errors by comparing clock estimates from Ka-band intersatellite links and L-band GNSS observations. For the L-band clock estimation, the 5-parameter ECOM was used, which is insufficient to consider the elongated shape of the MEO satellites. These orbit modeling deficiencies are reflected as a bump at roughly half of the orbital period in the Allan deviation. Ka-band clocks were estimated from geometry-free ISL observations not suffering from these systematic errors. \cite{Steigenberger_2015} showed similar effects for early Galileo orbit and clock products of four MGEX ACs: a clear dependence of the apparent clock stability on the elevation of the Sun above the orbital plane ($\beta$-angle) responsible for variations in the orbit quality. 
%These effects could be significantly reduced for Galileo by applying a more sophisticated orbit model, namely an a priori cuboid model \citep{Montenbruck_2015} or ECOM-2 \citep{Arnold_2015}.

Figure~\ref{fig:clockrms} shows the median clock RMS values of all \mbox{BeiDou-3} satellites covered by the MGEX ACs for the time period 66\,--\,365/2021 after fitting a 2nd order polynomial. The satellites are grouped by their block type and within each block sorted by their SVN. Gross outliers exceeding \SI{1}{ns} have been excluded. For SHAO, only a limited time period of November and December 2021 is considered. Earlier SHAO clock products suffer from an erroneous alignment to broadcast clocks introducing periodic variations that significantly degrade the clock quality. For the MEO satellites, CODE in general shows the smallest RMS values with \SIrange{70}{170}{ps} for CAST and  \SIrange{75}{205}{ps} for SECM. GFZ, IAC, and WUM are on a pretty similar level, whereas the RMS values for SHAO are increased for dedicated satellites, namely the CAST satellites C209, C210 and the SECM satellites C215, C225, and C226. However, due to the limited time interval for SHAO, these results have to be interpreted with care. For C222 and C223, also WUM shows increased RMS values of about \SI{250}{ns}. Most SECM MEO satellites have similar RMS values of around \SI{100}{ps} for all ACs except for C225. Although later CAST MEO as well as the SECM MEO satellites are equipped with PHM clocks, no clear performance advantage of that clock type compared to the RAFSs can be seen. %This might be related to orbit modeling deficiencies masking the performance of the physical clocks. 

The IGSO satellites show slightly increased RMS values of \SIrange{120}{250}{ps} compared to the MEOs. This might be partly attributed to the lower orbit quality (see Sec.~\ref{sec:orbPrecision}) as well as the fact that these satellites are only tracked by a regional network. The GEO satellites, that are only covered by a small set of ACs (see Table~\ref{tab:MGEXAC}), show the highest RMS values with \SIrange{275}{440}{ps}. Like for the IGSOs, this is attributed to the limited orbit quality caused by the static viewing geometry and the related correlations mentioned in Sec.~\ref{sec:orbPrecision}.

In addition to the orbit SISRE, Table~\ref{tab:sisre} contains overall SISRE values that also include the clock differences T in the SISRE computation:
\begin{equation}
    \text{SISRE}=\sqrt{w_1^2 R^2 - 2w_1RT + T^2 + w_2^2 \left(A^2+C^2 \right)} 
\end{equation}
In order to account for biases in the system time realization of the various ACs, an epoch-wise constellation mean clock value is removed. The SISRE of the MEO satellites varies between 3 and \SI{7}{cm}. For most combinations of ACs, the overall MEO SISRE is smaller than the orbit SISRE by several millimeters. This is possible due to a compensation of orbit-related errors by corresponding errors in the estimated satellite clocks. For the IGSO satellites, the best agreement is achieved for GFZ/WUM with \SI{8}{cm}, the worst for COD/SHA with \SI{20}{cm}. The overall IGSO SISRE is smaller than the orbit SISRE for all combinations of ACs with a maximum value of more than \SI{4}{cm}. The GEO SISRE values are on a level of \SIrange{5}{6}{dm} and almost identical to the orbit SISRE.

\begin{figure*}
    \centering
    \includegraphics[width=6cm]{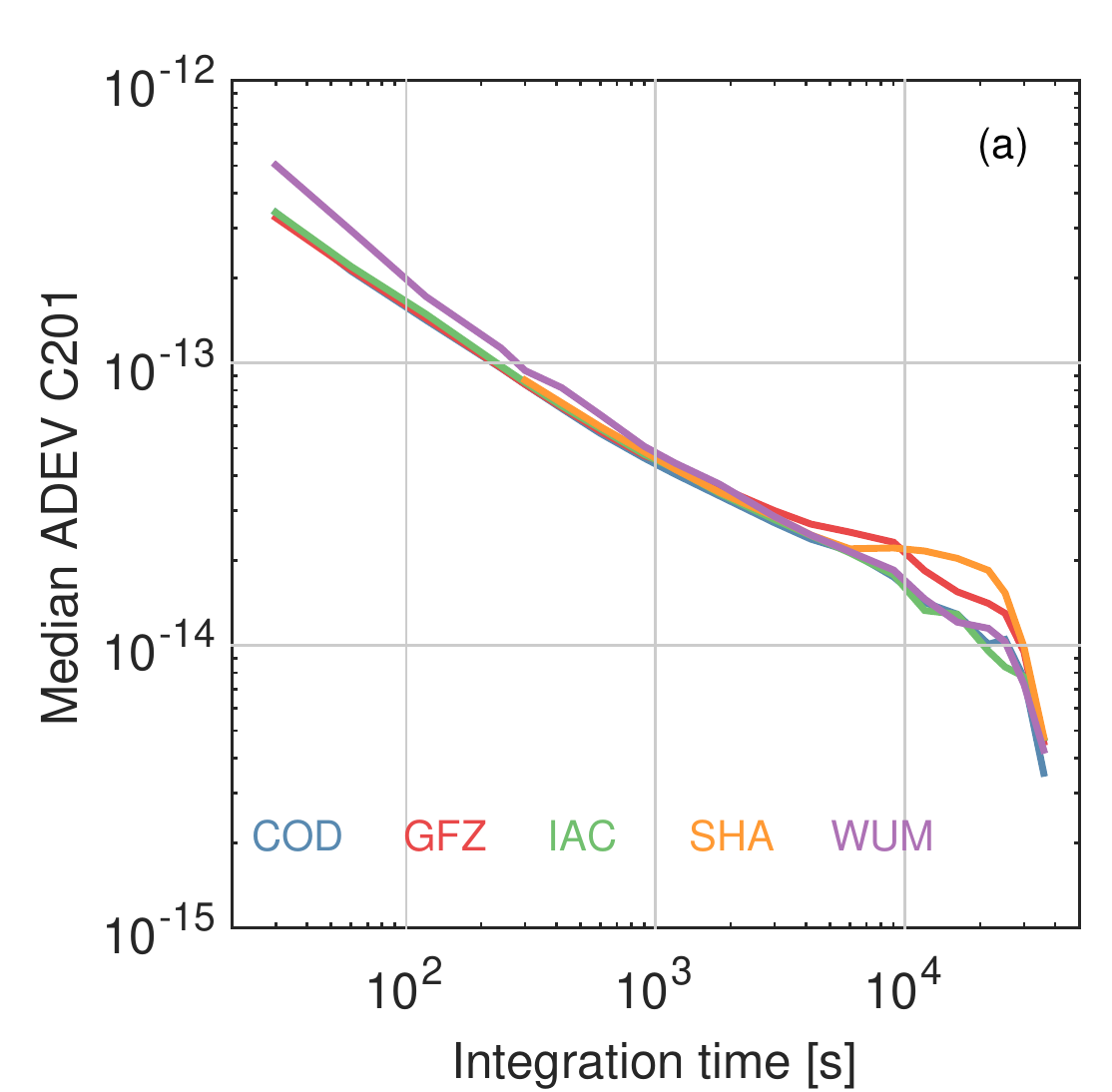} % C19
    \includegraphics[width=6cm]{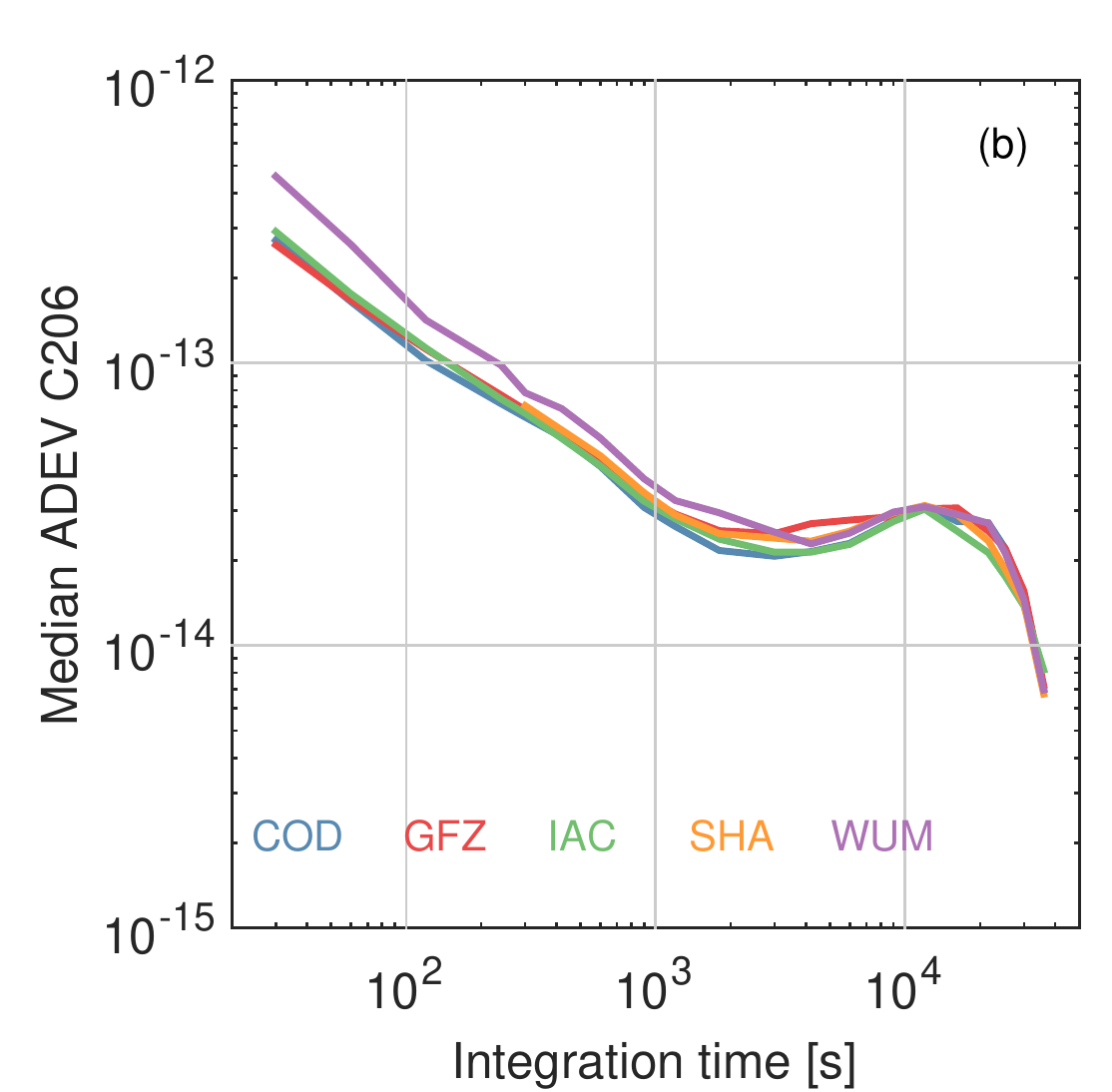} % C21
    \includegraphics[width=6cm]{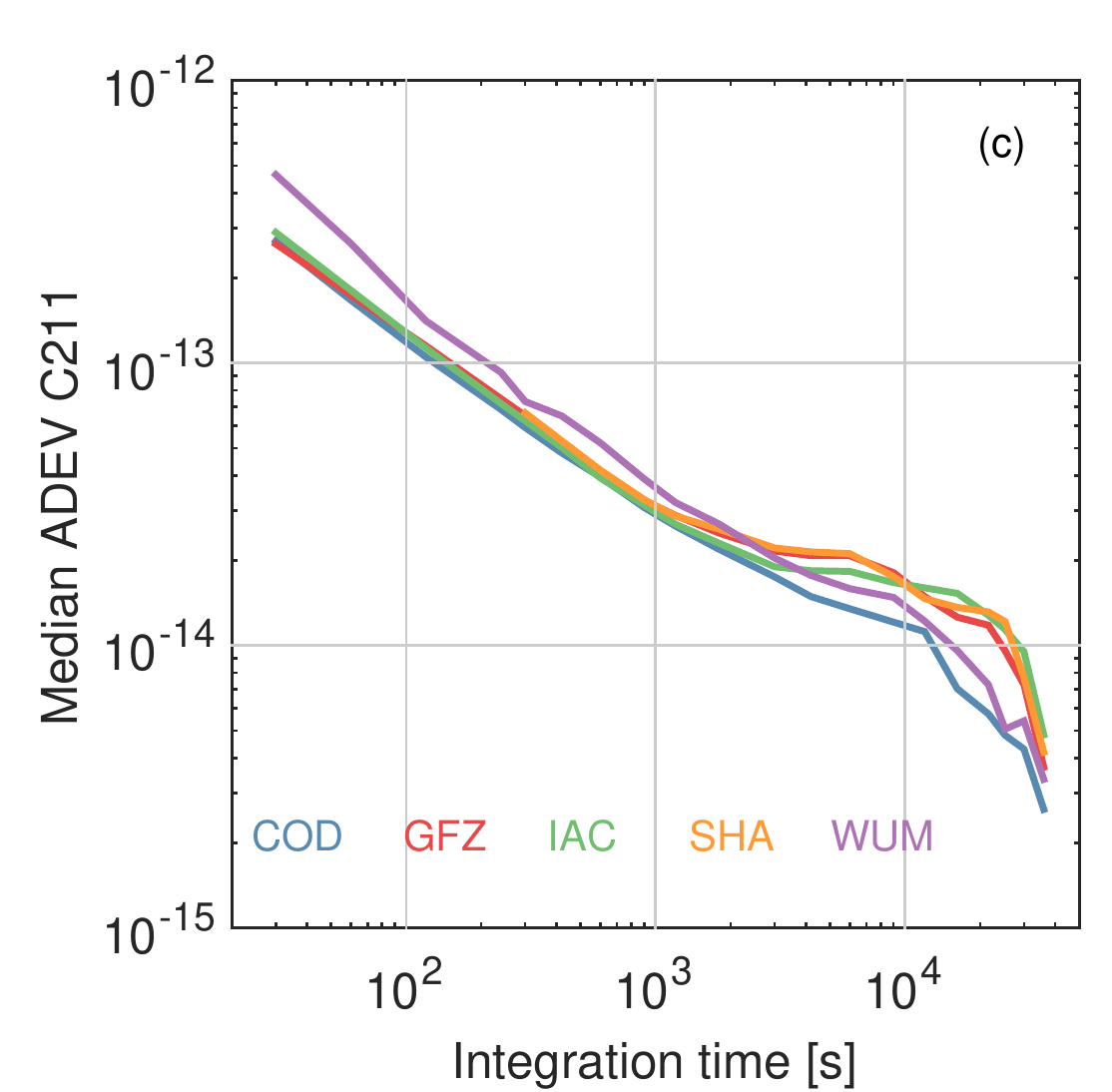} % C26
    \includegraphics[width=6cm]{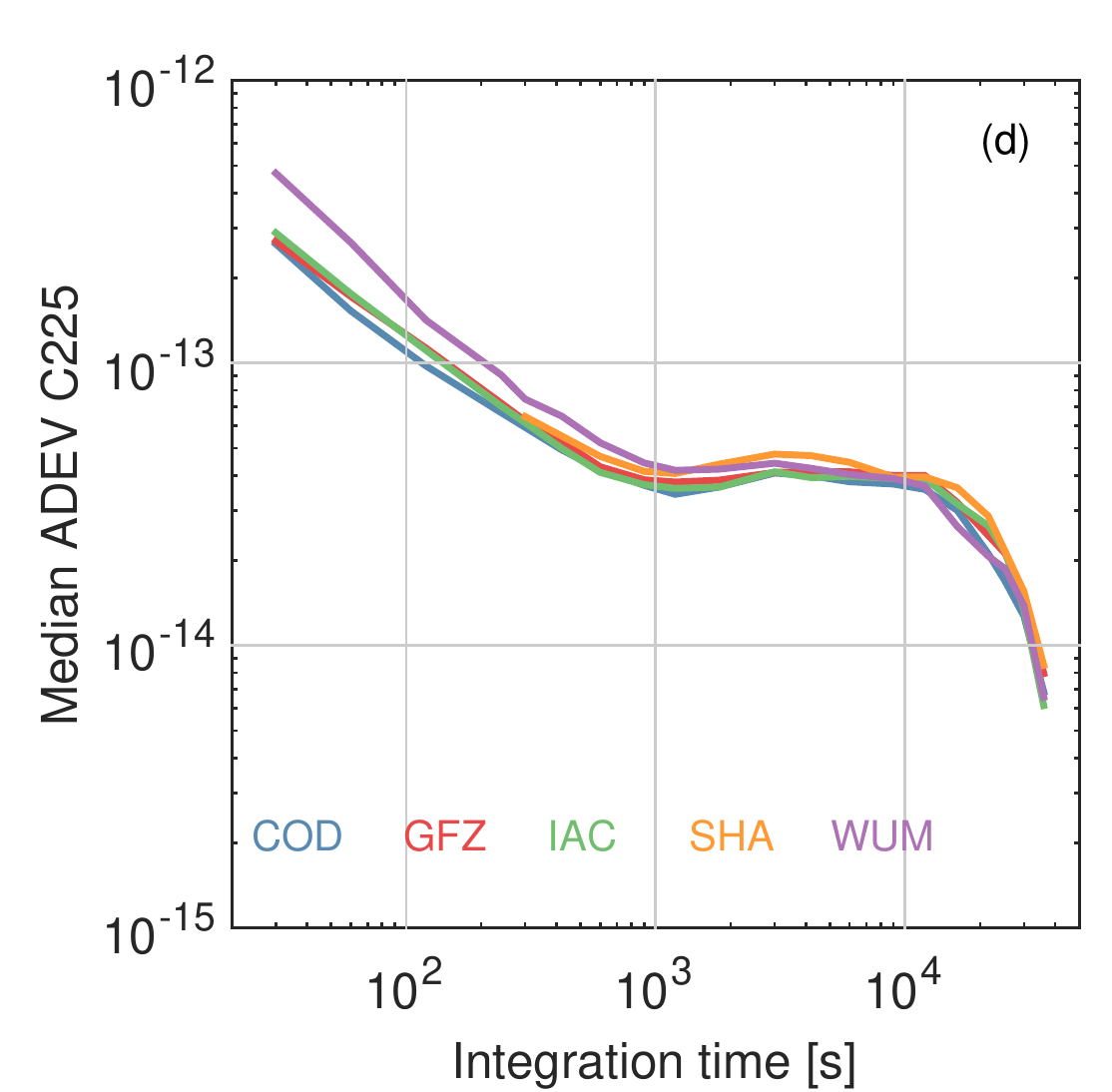} % C44
    \includegraphics[width=6cm]{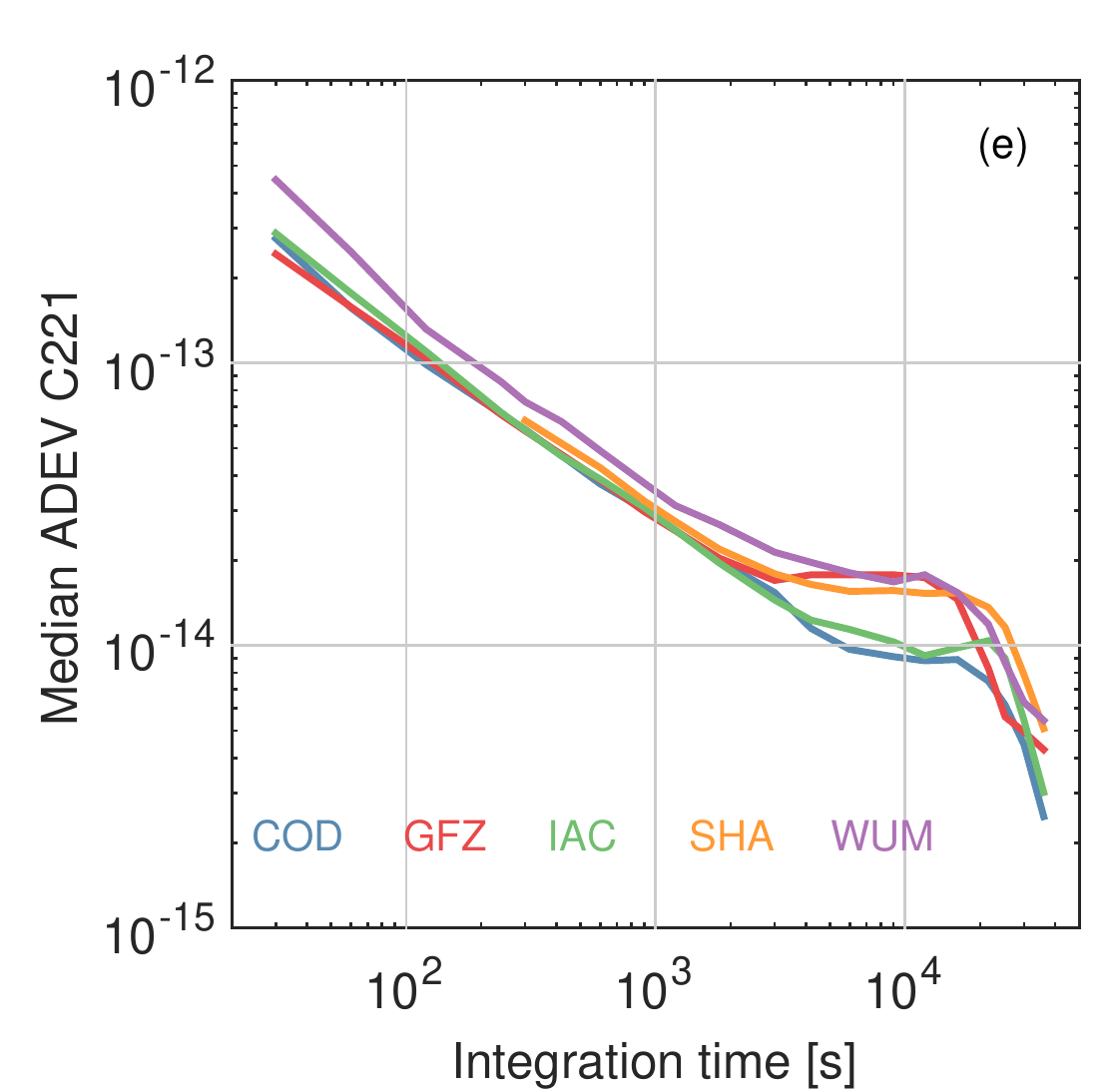} % C39 
    \includegraphics[width=6cm]{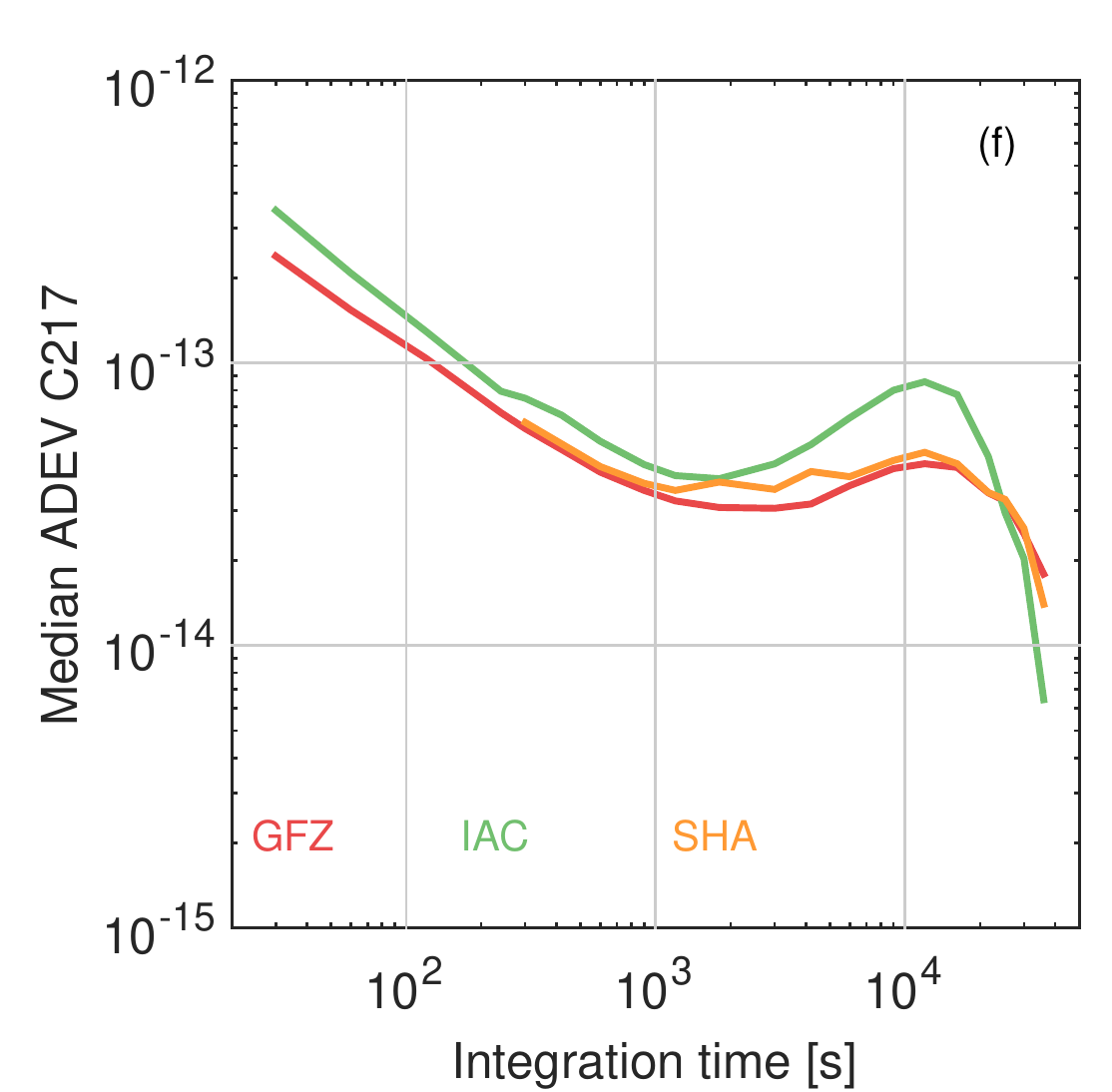} % C59
    \caption{Median Allen deviations for GPS Week 2188:  CAST MEO satellite C201 (a) and C206 (b) in orbital plane B with $\beta\approx$\SI{-41}{\degree}; SECM MEO satellites C211 (c) in orbital plane C with $\beta\approx$\SI{-34}{\degree} and C225 (d) in orbital plane A  with $\beta\approx$\SI{32}{\degree}; IGSO satellite C221 (e) with $\beta\approx$\SI{-78}{\degree}; GEO satellite C217 (f) with $\beta\approx$\SI{-24}{\degree}.}
    \label{fig:adev}
\end{figure*}
% Middle week 2188 => MJD 59563

Figure \ref{fig:adev} compares the apparent clock stability of selected satellites for the different ACs. Median Allan deviations (ADEVs) for GPS Week 2188 (December 12\,--\,18, 2021) are shown. SHAO provides \SI{5}{min} clock products whereas the other ACs have a higher sampling of \SI{30}{s}. Therefore, the ADEV plots for SHAO only start at an integration time of \SI{300}{s} whereas the other ACs start at \SI{30}{s}. For integration times up to \SI{1000}{s}, the ADEV values of WUM are slightly increased for all satellites compared to the other ACs. \cite{Gu_2021b} already reported slightly increased ADEV values of WUM compared to GFZ for integration times up to \SI{5000}{s}.

The finding of \cite{Qin_2020} that the \mbox{BDS-3} RAFSs and PHMs have a similar overall performance is generally confirmed by the anaylsis of the MGEX clock products. However, clock- or satellite-specific performance differences may still be noted, which deserve further investigation. By way of example, we consider the CAST MEO satellites C201 and C206 both operating on their RAFS. Both occupy the same orbital plane and their apparent clock stability should be equally affected by potential SRP modeling deficiencies. Nevertheless, both satellites exhibit clearly distinct ADEV characteristics. The RAFS of the CAST MEO satellite C201 in Fig.~\ref{fig:adev}a shows a steadily decreasing ADEV while the C206 RAFS (Fig.~\ref{fig:adev}b) shows a higher stability for integration times up to \SI{2000}{s} but also a significant bump at \SI{20000}{s}. Similar differences can be observed for the SECM satellites C211 (Fig.~\ref{fig:adev}c) and C225  (Fig.~\ref{fig:adev}d), which both use a PHM clock. The two spacecraft are placed in different orbital planes but the absolute values of their $\beta$-angles are quite similar. Nevertheless, C211 only shows a small bump at \SI{10000}{s}, whereas C225 has a kind of plateau between 1000 and \SI{10000}{s}. These findings suggest that on top of possible SRP modeling issues the apparent clock stability of individual satellites is affected by thermal variations across the orbit. These may either relate to frequency variations of the actual clock or bias variations in the transmitter chain but would likely require onboard telemetry for further information. 

The PHM of the IGSO satellite C221 (Fig.~\ref{fig:adev}e) shows a similar stability like the MEO satellites. CODE and IAC have both a slightly smaller ADEV for integration times around \SI{10000}{s}. However, the other two IGSO satellites C220 and C224 show bumps of $5\cdot10^{-14}$ at an integration time of \SI{10000}{s}. These differences are likely related to the different $\beta$-angles. C220 and C224 have $\beta$-angles of about 8 and \SI{13}{\degree}, respectively, whereas the increased $\beta\approx$\SI{-78}{\degree} for C221 is responsible for smaller orbit-induced clock errors. These errors affect the apparent clock stability most prominently for GEO satellites as illustrated in Fig.~\ref{fig:adev}f for the PHM of C217. The ADEV flattens at \SI{1000}{s} and reaches up to $8 \cdot 10^{-14}$ at \SI{15000}{s}.

\begin{figure*}
  \centering
  \includegraphics[width=\textwidth]{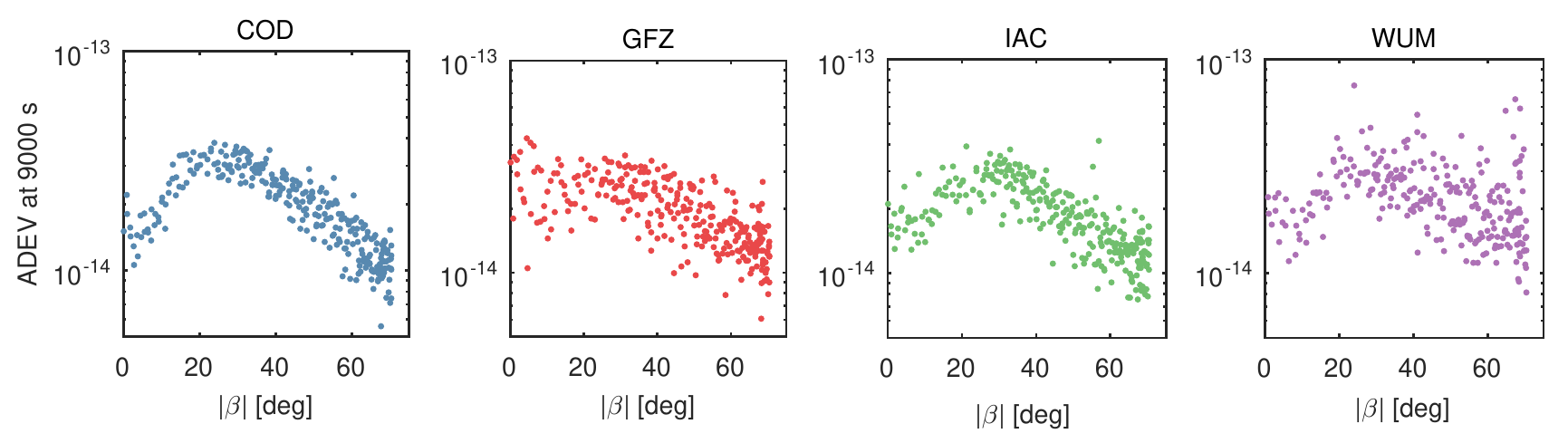}
  \caption{Allan deviation at \SI{9000}{s} integration time for C213 (CAST, RAFS) versus absolute value of the elevation of the Sun above the orbital plane.}
  \label{fig:adevBeta}
\end{figure*}

The ADEVs in Fig.~\ref{fig:adev} only provide a snapshot of the apparent clock stability for one week of data. Orbit mismodeling issues widely depend on the elevation of the Sun above the orbital plane. As radial orbit errors are one-to-one correlated with clock errors, the ADEV is expected to also show such systematic effect. By way of example, the ADEV value for \SI{9000}{s} integration is plotted against the absolute value of  $\beta$ in Fig.~\ref{fig:adevBeta} for the CAST MEO satellite C213 to demonstrate the $\beta$-dependence of the apparent clock stability. Due to the clock datum issues mentioned above, SHAO is not considered in that figure. All ACs reveal a clear $\beta$-dependence but the individual patterns differ. CODE shows maximum ADEV values of $5\cdot10^{-14}$ at \SIrange{20}{30}{\degree} and the smallest scatter of all ACs \wrt a $\beta$-dependent mean value. The ADEV values of GFZ have an almost linear $\beta$-dependency. IAC and WUM have in general a similar pattern with a maximum at about \SI{30}{\degree} but WUM has a significantly higher scatter. The different ADEV pattern might reflect the orbit modeling strategies of the different ACs: CODE is the only AC applying \mbox{ECOM-2} whereas GFZ uses ECOM and both ACs do not consider an a priori SRP model. IAC and WUM both rely on an a priori box-wing model but from different sources complemented by ECOM for WUM and an ECOM-based parameterization for IAC (see Tab.~\ref{tab:AC_opt}).

\begin{table}
    \centering
    \caption{Average 1-$\sigma$ repeatability  of the daily BDS-3 MEO (C) and GPS (G) PPP solutions in December 2021 over the ensemble of 30 stations.} 
    \begin{tabular}{lrrrrrrrr} \toprule
        AC  &  \multicolumn{2}{c}{North} & \multicolumn{2}{c}{East} & \multicolumn{2}{c}{Up }  & \multicolumn{2}{c}{3D }\\
                  &  \multicolumn{2}{c}{[mm]} & \multicolumn{2}{c}{[mm]} & \multicolumn{2}{c}{[mm]}  & \multicolumn{2}{c}{[mm]}\\
            &  C  &  G  &   C  &  G  &  C  &  G  &  C  &  G   \\ \midrule
        COD & 3.0 & 2.8 &  3.2 & 3.0 & 5.7 & 5.0 &  7.2 & 6.6  \\  
        GFZ & 2.1 & 2.1 &  2.6 & 2.5 & 5.5 & 5.1 &  6.6 & 6.1  \\ 
        IAC & 2.7 & 2.6 &  4.0 & 3.8 & 6.3 & 5.7 &  8.0 & 7.4  \\ 
        SHA & 4.1 & 2.2 &  7.1 & 3.5 &10.5 & 6.2 & 13.4 & 7.6  \\ 
        WUM & 2.7 & 2.3 &  3.0 & 2.7 & 6.2 & 5.2 &  7.5 &  6.4 \\ \bottomrule  
        
    \end{tabular}   
    \label{tab:PPP}
\end{table}

%%%%%%%%%%%%%%%%%%%%%%%%%%%%%%%%%%%%%%%%%%%%%%%%%%%%%%%%%%%%%%%%%%%%%%%%%%%%%
\section{Precise point positioning}                           \label{sec:ppp}
%%%%%%%%%%%%%%%%%%%%%%%%%%%%%%%%%%%%%%%%%%%%%%%%%%%%%%%%%%%%%%%%%%%%%%%%%%%%%
Precise point positioning \citep[PPP;][]{Zumberge_1997} directly utilizes precise orbit and clock products. Therefore, the PPP performance of the MGEX BDS-3 products allows for an indirect assessment of the orbit and clock quality. For this purpose, we computed static PPP solutions of 30 stations with the NAPEOS software \citep{Springer_2009a} on a daily basis for December 2021. The stations are equipped with different receiver and antenna types and globally distributed. The ionosphere-free linear combination of B1 and B3 observations is processed at a sampling interval of \SI{5}{min}. Only observations of BDS-3 MEO satellites are considered. In addition to station coordinates, troposphere zenith delays are estimated every two hours, receiver clocks every epoch and ambiguities are estimated as float parameters. The elevation cutoff angle is set to \SI{10}{\degree} and elevation-dependent weighting with $w=\sin\epsilon$ is applied. The receiver and satellite antenna models have been selected according to Table~\ref{tab:AC_opt}. For comparison purposes, also a GPS-only solution with L1 and L2 P(Y) observations has been computed.

The station-averaged repeatabilities of the daily position solutions are given in Table~\ref{tab:PPP} for each AC. The RMS of the North component for the \mbox{BDS-3} MEO PPP is on the average level of \SIrange{2}{4}{mm} with an overall range of \SIrange{2}{5}{mm}. The East component RMS is slightly higher for most ACs with values between 2 and \SI{11}{mm} for individual stations. Compared to the other ACs, SHA shows a slightly degraded performance for BDS-3 but is on the same level for GPS. Due to correlations between station heights, receiver clock, and troposphere parameters, the precision of the station heights is generally worse than for the horizontal components with average RMS values between 6 and \SI{10}{mm} and individual values between 4 and \SI{16}{mm}. The overall 3D RMS is on the level of \SIrange{7}{8}{mm} except for SHA and ranges from 4 to \SI{22}{mm} for individual stations. 

Compared to BDS-3 MEO PPP solutions, the GPS results have slightly smaller RMS values. This can probably be attributed to the increased number of satellites (31 vs.\ 24) resulting in an improved observation geometry. However, the differences are mostly on the level of a few tenth of a millimeter with a maximum of one millimeter when SHA is excluded. 

Inconsistencies of the models applied in orbit and clock product generation and the PPP can degrade the PPP results. Therefore, the results in Table~\ref{tab:PPP} are likely worse than PPP solutions generated with fully consistent modeling. However, as many users will utilize different software packages than those used for the product generation, the numbers in Table~\ref{tab:PPP} are considered to be representative for BDS-3 PPP performance from a user's perspective.

Another measure for the consistency of multi-GNSS orbit and clock products are coordinate differences obtained from the processing of different GNSSs. Table~\ref{tab:PPPdif} lists the mean coordinate differences between GPS-only and BDS-3 MEO PPP as well as their STD for the 30 stations already used before. On average over all stations, the horizontal coordinates based on GPS and BDS-3 MEOs match on the level of \SIrange{1}{2}{mm} with an RMS scatter of about \SI{6}{mm}. For the vertical component, which suffers from an inferior dilution of precision, the difference of GPS- and BDS-3-based positions exhibits mean differences between $-6$ and \SI{13}{mm} for the individual solutions with STDs of \SIrange{8}{16}{mm} across the 30 stations. The best consistency of the vertical coordinates between GPS and BDS-3 MEO solutions is obtained with the GFZ, IAC, and WUM products. A slightly degraded performance is obtained for COD which is surprising as COD uses dedicated B1/B3 receiver antenna calibrations whereas the other ACs apply the GPS L1/L2 calibrations also for BDS B1/B3. An increased scatter is also observed for SHA, which is subject to further investigation.

\begin{table}
  \caption{Mean and standard deviation of of monthly-averaged coordinate differences between GPS and BDS-3 PPP solutions in December 2021 over the ensemble of 30 stations.} 
  \centering
  \begin{tabular}{lrrrrrr} \toprule
  AC  & \multicolumn{2}{c}{North [mm]} & \multicolumn{2}{c}{East [mm]} &  \multicolumn{2}{c}{Up [mm]} \\ 
      & Mean&  STD &   Mean &  STD &  Mean  &  STD \\ \midrule
  COD &   0.7 &  2.9 & $-$2.3 &  4.7 & $-$6.1 & 11.7 \\
  GFZ &   0.6 &  2.3 & $-$0.7 &  3.8 & $-$0.9 &  7.9 \\
  IAC &   0.3 &  2.0 & $-$0.8 &  4.2 & $-$4.7 &  8.1 \\
  SHA &$-$0.6 &  3.1 & $-$0.2 &  6.4 &    6.4 & 10.6 \\
  WUM &   0.4 &  1.9 &    0.1 &  4.5 & $-$4.6 &  8.4 \\ \bottomrule
  \end{tabular}
  \label{tab:PPPdif}
\end{table}

%%%%%%%%%%%%%%%%%%%%%%%%%%%%%%%%%%%%%%%%%%%%%%%%%%%%%%%%%%%%%%%%%%%%%%%%%%%%%
\section{Summary and outlook}
%%%%%%%%%%%%%%%%%%%%%%%%%%%%%%%%%%%%%%%%%%%%%%%%%%%%%%%%%%%%%%%%%%%%%%%%%%%%%
BeiDou-3 officially provides operational global positioning and timing services since July 2020. Wuhan University was the first MGEX AC including BDS-3 in their orbit and clock products in January 2019. As of beginning of 2022, five MGEX ACs include BeiDou-3 in their operational products. The orbit consistency of the five MGEX ACs is on the level of \SIrange{4}{8}{cm} 3D RMS for the MEO satellites, \SIrange{10}{20}{cm} for the IGSO satellites and about \SI{60}{cm} for the GEO satellites. The orbit accuracy of the MEO satellites as evaluated by SLR residuals is on the level of \SIrange{4}{9}{cm}. However, the current SLR tracking of BeiDou-3 is limited to only four MEO satellites, while no IGSO or GEO satellites are tracked at all. In order to better assess the accuracy of the BeiDou-3 orbits and to understand potential modeling deficiencies of individual satellites, an extended coverage of the SLR tracking of the BeiDou-3 constellation is therefore strongly encouraged.

For the BDS-3 MEO satellites, the MGEX clock products show a similar performance as regards median clock RMS values. They vary between 75 and \SI{250}{ps}. Differences for individual satellites are consistently visible for all ACs but no general difference for RAFS and PHM clocks can be identified. The IGSO satellites have slightly higher RMS values of up to \SI{260}{ps} whereas the GEO satellites reach up to \SI{440}{ps}. However, it has to be mentioned that these values refer to the apparent clock stability degraded by SRP modeling deficiencies. This is confirmed by $\beta$-dependent patterns in the Allan deviations. These are visible for all ACs but have differences in shape due to different SRP modeling strategies. 

The combined orbit and clock precision as evaluated by the signal-in-space range error for pairs of ACs is between 3 and \SI{7}{cm} for the MEO satellites, on the one to two decimeter level for the IGSO satellites and reaches \SI{6}{dm} for the GEO satellites. Precise point positioning allows for an assessment of the MGEX orbit and clock products in the position domain. The precision of the horizontal components is on the few millimeter level except for SHAO, while the height component varies between 6 and \SI{8}{mm}. However, systematic differences between GPS- and BDS-3-derived coordinates point to modeling deficiencies degrading the overall accuracy. It remains to be investigated how use of the modernized B1C and B2a signals of BeiDou-3 and a treatment of  BDS-2 and -3 as independent constellations can help to further improve the consistency of GPS and BeiDou processing.   

The orbit and clock quality of BeiDou-3 does not yet reach that of GPS or Galileo. However, this is not astonishing as the IGS ACs have more than three decades of experience with GPS data processing and one decade for Galileo whereas the first BDS-3 satellite was only launched in 2017. A way forward to improve the orbit quality could be the release of more detailed metadata. As regards the optical properties, only the absorption coefficients are published whereas specular and diffuse reflection coefficients are missing. \cite{Duan_2021d} pointed to the presence of a search and rescue (SAR) antenna on several MEO satellites but no information about its geometry as well as the optical properties are available. Furthermore, information on the size and optical properties of the communication antennas of the IGSO and GEO satellites could also help to improve the orbit quality of these types of satellites. 

The ultimate goal of the MGEX pilot project is a combined multi-GNSS orbit and clock product covering all constellations. First steps towards a combined orbit product were conducted by GFZ \citep{Mansur_2020a,Mansur_2020b} and Geoscience Australia \citep{Sosnica_2020}. \cite{Zhou_2022b} present orbit combination results of ACs of the international GNSS Monitoring and Assessment System (iGMAS). They report a \SIrange{2}{3}{cm} consistency evaluated by 1D RMS values of the AC contributions \wrt the combined orbits. The corresponding clock consistency is about \SI{20}{ps}. The SLR biases are around \SI{+1}{cm} for the CAST and \SI{-5}{cm} for the SECM satellites with STDs between 3 and \SI{4}{cm}. 

One critical issue when combining \mbox{BDS-3} satellite orbits are the scale differences between the various ACs as discussed in Sect.~\ref{sec:orbAccuracy}. Whereas parts of these discrepancies might be reduced by a more advanced or at least a more consistent orbit modeling, the remaining differences degrade the quality of the combination and one might think of correcting/adjusting them before the combination. 

While there is obviously room for further improvement of the precision and accuracy of present \mbox{BDS-3} MGEX products, it is evident that \mbox{BDS-3} can already make valuable contributions to geodesy and precise point positioning. As a fully global system, it provides a highly valuable complement of the well-established GPS and other GNSSs. Continued effort by all IGS ACs will help to better characterize the \mbox{BeiDou-3} constellation and to remove remaining deficiencies in the orbit dynamics and measurement modeling. This will help to further harmonize the individual AC solutions and finally enable generation of a combined multi-GNSS orbit and product of utmost performance.

\section*{Acknowledgements}
We'd like to acknowledge the efforts of the station operators and data centers supporting IGS MGEX. The MGEX activities at Shanghai Observatory are supported by the National Natural Science Foundation of China (12073063). The MGEX activities at Wuhan University are supported by the National Natural Science Foundation of China (41974035) and the Young Elite Scientists Sponsorship Program by China Association for Science and Technology (2018QNRC001).

\section*{Data Availability}
IGS MGEX orbit and clock product files \citep{IGS_2021} are available at selected global IGS data centers: Crustal Dynamics Data Information System (CDDIS, USA): \url{https://cddis.nasa.gov/archive/gnss/products/mgex/} and Institut National de l'Information Géographique et Forestière (IGN, France): \url{ftp://igs.ign.fr/pub/igs/products/mgex/}.

%% Bibliography
%% Author year style
\bibliographystyle{model5-names}
\biboptions{authoryear}
\bibliography{bds3_mgex}
    
\end{document}